\documentclass{eas}
\usepackage{graphicx}
\usepackage{natbib}
\usepackage{amsmath}
\usepackage{amssymb}
\newcommand{\der}{\ensuremath{\,\mathrm{d}}}
\newcommand{\dv}{\ensuremath{\,\mathrm{d}v}}

\newcommand{\dnu}{\ensuremath{\,\mathrm{d}\nu}}
\newcommand{\domega}{\ensuremath{\,\mathrm{d}\omega}}
\newcommand{\pul}{\ensuremath{\frac{1}{2}}}
\newcommand{\japul}{\ensuremath{\frac{3}{2}}}

\newcommand{\zav}[1]{\left(#1\right)}
\newcommand{\hzav}[1]{\left[#1\right]}

\newcommand{\deriv}[2]{\frac{{\mathrm d}#1}{{\mathrm d}#2}}
\newcommand{\mn}{\ensuremath{{\mu\nu}}}

\newcommand{\NL}{\ensuremath{\mathrm{NL}}}
\newcommand{\NLtot}{\ensuremath{\mathrm{NL}_\mathrm{tot}}}

\defcitealias{ja1}{these proceedings}
\defcitealias{nicenp}{these proceedings}
\defcitealias{nicekp}{these proceedings}
\defcitealias{nicelm}{these proceedings}
\TitreGlobal{Non-LTE Line Formation for Trace Elements in Stellar
Atmospheres}
\begin{document}

%%-----------------------------
%%      the top matter
%%-----------------------------
\title{Statistical equilibrium equations for trace elements in stellar
atmospheres}
\runningtitle{Statistical equilibrium equations for trace elements}
\author{Ji\v{r}\'{\i} Kub\'at}\address{Astronomick\'y \'ustav AV \v{C}R,
Fri\v{c}ova 298, 251 65 Ond\v{r}ejov, Czech Republic}

\begin{abstract}
The conditions of thermodynamic equilibrium, local thermodynamic
equilibrium, and statistical equilibrium are discussed in detail.
The equations of statistical equilibrium and the supplementary equations
are shown together with the expressions for radiative and collisional
rates with the emphasize on the solution for trace elements.
\end{abstract}

\maketitle

%%%%%%%%%%%%%%%%%%%%%%%%%%%%%%%%%%%%%%%%%%%%%%%%%%%%%%%%%%%%%%%%%%%%%%%%
\section{Introduction: Stellar atmosphere problem}

Solution of the stellar atmosphere problem is usually considered as the
determination of spatial dependence of basic macroscopic quantities
\citep[see also][\citetalias{ja1}]{ja1} also be understood
\citep[following][]{ivandis} as aiming at finding three basic
microscopic distributions, namely, the momentum distribution
(distribution of velocities of all particles), distribution of particle
internal degrees of freedom (populations of atomic excitation stages),
and distribution of internal degrees of freedom of the electromagnetic
field (radiation field for all frequencies and directions).

%%%%%%%%%%%%%%%%%%%%%%%%%%%%%%%%%%%%%%%%%%%%%%%%%%%%%%%%%%%%%%%%%%%%%%%%
\section{From thermodynamic equilibrium to NLTE}

%=======================================================================
\subsection{Thermodynamic equilibrium}

To maintain thermodynamic equilibrium, several necessary conditions have
to be met (see \citealt{ivandis}, also \citealt{ivanlnp}).
First, the characteristic time for macroscopic changes of the medium has
to be much larger than the relaxation time
($t_\mathrm{relaxation} \ll t_\mathrm{macroscopic\ changes}$).
Second, macroscopic changes of the medium must be small on the scale of
a particle mean free path
($l_\mathrm{macroscopic\ changes} \ll \bar l_\mathrm{free\ path}$).
Then, the relaxation time has to be much less than the time interval
between inelastic collisions
($t_\mathrm{relaxation} \ll t_\mathrm{inelastic\ collisions}$).
And, finally, if the latter condition is violated, then colliding
particles need to have equilibrium distributions.

If the medium is in thermodynamic equilibrium, then all three
distributions mentioned in Introduction (momentum, internal, and
radiation) have their equilibrium form.
Velocities $v_\mathrm{a}$ of particles $\mathrm{a}$ follow the
Maxwellian distribution,
\begin{equation}\label{jk3maxw}
f(v_\mathrm{a}) \dv_\mathrm{a}= \frac{1}{\bar{v}_\mathrm{a}
\sqrt{\pi}}e^{-\zav{{v_\mathrm{a}^2}/{\bar{v}_\mathrm{a}^2}}}
\dv_\mathrm{a},
\end{equation}
where $\bar{v}_\mathrm{a} = \sqrt{{2kT_\mathrm{a}}/{m_\mathrm{a}}}$
is their most probable speed, $T_\mathrm{a}$ is the corresponding particle
temperature, $m_\mathrm{a}$ is its mass, and $k$ is the Boltzmann
constant.
In thermodynamic equilibrium all temperatures $T_\mathrm{a}$ are the
same and in the following we denote them as $T$.

Individual atomic energy levels $i$ are populated according to the
Boltzmann distribution \citep[see][Eq.\,5-4]{mih},
\begin{equation}\label{jk3boltz}
\frac{n_i^*}{n_0^*} = \frac{g_i}{g_0} e^{-\frac{\chi_i}{kT}},
\end{equation}
where $n_i$ is the population of the $i$-th level, $n_0$ is the
population of the ground level, asterisks ($*$) denote equilibrium
values, $g_i$ is the statistical weight of the level $i$, and $\chi_i$
is its excitation energy.
The ionization equilibrium is described by the Saha equation
\citep[see][Eq.\,5-15]{mih}
\begin{equation}\label{jk3saha}
\frac{N^*_j}{N^*_{j+1}} = n_e \frac{U_j(T)}{2U_{j+1}(T)}
\zav{\frac{h^2}{2\pi m_e kT}}^\japul e^{\frac{\chi_{Ij}}{kT}},
\end{equation}
where $N_j^\ast$ is the total population (number density) of the ion
$j$, $U_j(T)$ is its partition function, $n_e$ is the electron number
density, $\chi_{Ij}$ is the ground state ionization energy, and $h$ is
the Planck constant.

Finally, the radiation field energy is distributed in frequency $\nu$
according to the Planck law
\begin{equation}\label{jk3planck}
B_\nu (T) = \frac{2h\nu}{c^2} \frac{1}{e^\frac{h\nu}{kT}-1}.
\end{equation}

%=======================================================================
\subsection{Local thermodynamic equilibrium (LTE)}

The fact that we can observe stellar light and the shape of the stellar
spectrum clearly show that the stellar radiation is far from that of a
blackbody.
Therefore, the assumption of radiation field energy equilibrium is not
valid and we have to include detailed radiative transfer.
However, simple relaxing of the equilibrium condition for radiation does
not help, since if the first two distributions are in equilibrium in a
homogeneous medium, then also the radiation field is forced to have the
equilibrium value.

To allow the light to escape from the star, we have to allow spatial
dependence of temperature and density.
We assume \emph{local} equilibrium distributions of particle velocities
\eqref{jk3maxw} and level populations (\ref{jk3boltz} and
\ref{jk3saha}), which means that the equilibrium condition ignores
possible gradients of $T(\vec{r})$ and $N(\vec{r})$.
Then, the equilibrium conditions change from point to point.
This approximation is usually called local thermodynamic equilibrium
(LTE).

To determine the specific radiation field intensity ($I_\mn$)
distribution, we solve the radiative transfer equation.
For the simple case of a plane-parallel horizontally homogeneous
atmosphere (1D case with the only spatial coordinate $z$), the transfer
equation along the ray determined by $\mu=\cos\theta$, where $\theta$
\citep[see Fig.\,1 in][\citetalias{ja1}]{ja1} is the ray angle with the
coordinate axis $z$, reads
\begin{equation}\label{jk3rtepp}
\mu \deriv{I_\mn}{z}
= \eta_\nu - \chi_\nu I_\mn,
\end{equation}
where $\eta_\nu$ and $\chi_\nu$ are the emissivity and opacity,
respectively.
The condition of local thermodynamic equilibrium implies that
$\eta_\nu = \chi_\nu B_\nu(T(z))$ at each depth $z$.

%=======================================================================
\subsection{Statistical equilibrium (NLTE)}

The non-equilibrium distribution of the radiation field obtained by
solution of \eqref{jk3rtepp} affects in turn the atomic level population
distribution.
If this influence becomes strong, the Saha-Boltzmann ionization and
excitation distribution is no longer in equilibrium.
But even if this influence is weak, it is better to assume that the
excitation distribution can generally be a non-equilibrium one.

Doing this, we assume that the excitation and ionization states
distribution does not follow from the Saha \eqref{jk3saha} and Boltzmann
\eqref{jk3boltz} equations.
For its determination we have to explicitly consider all individual
level populating and de\-populating processes.
Such treatment is usually referred to as NLTE or non-LTE.

Assuming NLTE we still assume an \emph{equilibrium} (i.e. Maxwellian -
Eq.~\ref{jk3maxw}) distribution of particle velocities, especially of
electrons.
The radiation field and level populations are allowed to have
non-equilibrium distributions, which for the radiation field is
determined by solution of the radiative transfer equation
\eqref{jk3rtepp} and for the level populations by solution of
statistical equilibrium equations \eqref{jk3ese} described later in the
Section~\ref{jk3secese}.

%-----------------------------------------------------------------------
\subsubsection{Microscopic processes}

Due to the non-equilibrium nature of level population distribution it is
necessary to take into account all processes influencing level
populations on a microscopic level.
The most important processes for hot star atmospheres are listed in
Table~\ref{mikroprocesy}.

\begin{table}
\caption{List of most important microscopic processes affecting the
equilibria in stellar atmospheres.
In this table, $\nu$ means photon, $\mathrm{e}$ stands for electron,
$\mathrm{X}$ means a particular atom.
Superscript $^+$ denotes the ion and $^*$ the excited state.
E (in the rightmost column) means that the process tends to establish
equilibrium of population numbers, M denotes processes that tend to
establish equilibrium velocity distributions.}
\label{mikroprocesy}
\label{rovnovahy}
\vspace{5pt}
\begin{tabular}{llllr}
\hline
\multicolumn{2}{l}{collisions} \\ & elastic &
\multicolumn{2}{l}{(e--e, e--H, e--H$^\text{\normalsize +}$, e--He,
H--H, H--He, ...)} & M \\
\cline{2-5}
& \multicolumn{2}{l}{inelastic with electrons} & & \\
& & excitation & $\mathrm{e}(v) + \mathrm{X} \rightarrow 
			\mathrm{e}(v^\prime < v) + \mathrm{X}^\ast$ &E\\
& & de-excitation & $\mathrm{e}(v) + \mathrm{X}^\ast \rightarrow
			\mathrm{e}(v^\prime > v) + \mathrm{X}$ &E \\
& & ionization & $\mathrm{e}(v) + \mathrm{X} \rightarrow
			\mathrm{e}(v^\prime<v) + \mathrm{e}
			(v^{\prime\prime}) + \mathrm{X}^+$ &E \\
& & recombination & $\mathrm{e}+ \mathrm{e}(v) + \mathrm{X}^+
			\rightarrow \mathrm{e}(v^\prime\ne v) +
			\mathrm{X}$ & E \\
\cline{2-5}
& \multicolumn{2}{l}{inelastic with other particles} & less frequent
	$\Rightarrow$ neglected & E\\
\hline
\end{tabular}
\\
\begin{tabular}{llllr}
\hline
\multicolumn{3}{l}{interaction with radiation} \\
& excitation & & $\nu + \mathrm{X} \rightarrow \mathrm{X}^\ast$ \\
\cline{2-5}
& de-excitation & spontaneous &
		$\mathrm{X}^\ast \rightarrow \nu + \mathrm{X}$ \\
& & stimulated & $\nu + \mathrm{X}^\ast \rightarrow 2\nu + \mathrm{X}$
\\
\cline{2-5}
& ionization & & $\nu + \mathrm{X} \rightarrow \mathrm{X}^+ + \mathrm{e}$
								\\
& & autoionization & $\nu + \mathrm{X} \rightarrow\mathrm{X}^{\ast\ast}
			\rightarrow \mathrm{X}^+ + \mathrm{e}$ \\
& & Auger ionization & $\nu +\mathrm{X} \rightarrow
			\mathrm{X}^{+\ast} + \mathrm{e}$ \\
\cline{2-5}
& recombination & spontaneous & $\mathrm{e} + \mathrm{X}^+
				\rightarrow \nu + \mathrm{X}$ & E \\
& & stimulated & $\nu + \mathrm{e} + \mathrm{X}^+ \rightarrow 2\nu +
				\mathrm{X}$ & E \\
& & dielectronic recombination & $\mathrm{X}^+ + \mathrm{e}
	\rightarrow \mathrm{X}^{\ast\ast} \rightarrow \nu + \mathrm{X}$
	&E \\
\hline
\end{tabular}
\\
\begin{tabular}{llll}
free-free transitions & & $\nu + \mathrm{e}+ \mathrm{X}
			\leftrightarrow \mathrm{e}+ \mathrm{X}$ & E\\
\hline
electron scattering & free (Compton, Thomson) &
	$\nu + \mathrm{e} \rightarrow \nu + \mathrm{e}$ & \\
& bound (Rayleigh) &
	$\nu + \mathrm{X} \rightarrow \nu + \mathrm{X}$ & \\
\hline
\end{tabular}
\end{table}

%=======================================================================
\subsection{Between LTE and NLTE}

%-----------------------------------------------------------------------
\subsubsection{Elastic collisions}

In both LTE and NLTE, a Maxwellian velocity distribution of particles
\eqref{jk3maxw} is assumed however it is never explicitly stated.
Elastic collisions between particles tend to maintain their equilibrium
velocity distribution.
These collisions are rarely explicitly considered both in LTE and in
NLTE calculations.
Usually, it is sufficient to assume that they are frequent enough
to balance all non-equilibrium processes that tend to destroy the
Maxwellian velocity distribution.
For stellar atmospheres, the equilibrium velocity distribution is a
reasonable assumption \citep[see][]{mih}.

%-----------------------------------------------------------------------
\subsubsection{Inelastic collisions}

Inelastic collisions play a key equilibrium role in the atomic level
populations.
If the velocity distribution is in equilibrium, then the inelastic
collisions push the atomic level populations also to their equilibrium.
On the other hand, inelastic collisions change the electron kinetic
energy and so cause departures from the Maxwellian velocity
distributions.
Equilibrium velocity distribution is recovered only if there is enough
elastic collisions.
Luckily, in most astrophysical cases the time between two inelastic
collisions is much larger than the time necessary for rebuilding the
equilibrium velocity distribution -- the relaxation time (the inequality
$t_\mathrm{relaxation} \ll t_\mathrm{inelastic\ collisions}$ holds), so
we can safely adopt Maxwellian velocity distribution.
However, there are exceptions, for example there is
evidence that the
electron velocity distribution differs from the Maxwellian
in the
solar transition region \citep{shoub} and in the solar corona
\citep{cranmer}.
In such cases we would have to solve an additional kinetic equation,
namely the equation for electrons.

%-----------------------------------------------------------------------
\subsubsection{Level populations}

In the following, we assume a Maxwellian velocity distribution for all
particles.
The distribution of the radiation field is, on the other hand, not in
equilibrium and must be determined by solution of the radiative transfer
equation.
Whether level populations will have their equilibrium value or not (in
other words, if the atmosphere will be ``LTE'' or ``NLTE'') depends on
the balance between processes connected with the equilibrium part --
inelastic collisions with electrons having Maxwellian distribution --
and processes connected with the non-equilibrium part -- radiative
processes, which connect the level populations with a generally
non-equilibrium radiation field.
In Table~\ref{rovnovahy} the letter (E) denotes processes that aim at
maintaining equilibrium of level populations.

If the equilibrium processes dominate (e.g. in a collisionally dominated
atmosphere), then level populations will have or will be close to the
equilibrium Saha-Boltzmann values.
In such case, all processes are in detailed balance, which means that
the rate of each process is balanced by the rate of the reverse process.
For a Maxwellian velocity distribution, the collisional processes are 
always in detailed balance.
Note that all radiative processes involving a free electron
(recombination, free-free transition) are in principle collisional.
On the other hand, radiative processes can hardly be in detailed
balance, they may be in detailed balance only for the case when
radiation has its equilibrium, i.e. Planck, distribution,
which happens only in the deepest parts of stellar atmospheres.
When the radiation intensity is different from Planckian, $J_\nu \ne
B_\nu$, then LTE is not an acceptable approximation.

If the non-equilibrium processes dominate, then a quite complicated game
begins.
The medium may reach states very far from equilibrium.

%%%%%%%%%%%%%%%%%%%%%%%%%%%%%%%%%%%%%%%%%%%%%%%%%%%%%%%%%%%%%%%%%%%%%%%%
\section{Trace elements}

Solution of the NLTE problem for trace elements may be considered
as a simplification of the standard model atmosphere problem, where we
aim to determine the temperature and density structures of the
atmosphere \citep[see][\citetalias{ja1}]{ja1}.
Here we assume that we have a given model atmosphere (either LTE or
NLTE), i.e. given temperature ($T(r)$) and density ($n_e(r)$)
structures, given chemical composition, and
if the given model atmosphere is a NLTE one, also
occupation numbers of atomic levels for the most important atmosphere
constituents.
This implies that also the background opacities (which are calculated
from the given model atmosphere) are known.
Then we solve simultaneously the radiative transfer and statistical
equilibrium equations for a trace element (or trace elements).

There are several conditions, which have to be met to enable the element
to be a trace element.
The basic requirement is that the influence of the trace element on the
atmospheric structure is really negligible.
The trace element may, of course, have influence on emergent radiation
in some parts of the spectrum.
More exactly, there is a requirement that inclusion of the NLTE trace
element does not change the atmospheric structure, which may also allow
to include the case of a change from an LTE treatment of the element to
the NLTE one.
This may sometimes enable to consider even hydrogen as a trace element
\citep[see also][\citetalias{nicenp}]{nicenp}, but this is not generally
recommended.
Stellar atmospheric structure is usually sensitive to changes in the
dominant ions, even if such changes are relatively small.
Sometimes it is very difficult to predict the final effect caused by a
change of the basic assumption from LTE to NLTE, so after any NLTE trace
element calculation it is necessary to check if the assumptions we made
at the beginning are still met.

This also means that we have always to check if a trace element is
really a trace element.
For example, changes in ionization equilibrium alter also the free
electron density, which may have an impact on the ionization structure
of the initial model atmosphere.
Background opacities have to be the same in the solution of the trace
element as in the model atmosphere.

There is one principal warning for using background LTE models, which is
based on physical grounds.
As a commonly used case we have an LTE model atmosphere, it means that
we assume that in the whole atmosphere there are sufficient collisions
with electrons to maintain equilibrium values of population numbers.
If we then choose one of elements as a trace element and solve the NLTE
problem, we implicitly say that the equilibrium may be violated for this
particular element.
But since the number of free electrons is the same as for all other
elements (which populations have their LTE values thanks to the
assumption of the LTE background model atmosphere), we simultaneously
assume that for the same physical conditions there exists detailed
balance for all transitions that belong to non-trace elements, but at
the same time it is lacking for the trace element.
Thus we are using two contradictory physical assumptions for the same
region.
Since LTE is a special case of a more general NLTE case, there may be
regions in the atmosphere, where this very specific condition is
fulfilled, but it hardly happens for the \emph{whole} atmosphere.
Using an LTE model atmosphere and assuming NLTE for a trace element is
highly inconsistent and one has always to prefer NLTE model atmospheres
as background models.

%%%%%%%%%%%%%%%%%%%%%%%%%%%%%%%%%%%%%%%%%%%%%%%%%%%%%%%%%%%%%%%%%%%%%%%%
\section{Equations of statistical equilibrium}\label{jk3secese}

For the change of the occupation number $n_i$  of the state $i$ of each
element in the stellar atmosphere we may write \citep[][Eq.\,5-48]{mih}
\begin{equation}\label{jk3eset}
\frac{\partial n_i}{\partial t} + \nabla \cdot \left(n_i \vec{v} \right)
= \sum_{\substack{l=1\\ l\ne i}}^{\NLtot}
\left(n_l P_{li} - n_i P_{il} \right),
\end{equation}
where $\NLtot$ is the total number of energy levels considered,
$P_{il}$ is the probability (per time unit) of a transition from
the state (level) $i$ to the state (level) $l$.
This probability consist of two parts, namely of the probability that
the transition is caused by absorption or emission of radiation
(the radiative probability $R_{il}$), or that the transition is caused
by a collision with a neighbouring particle, mostly with an electron
(the collisional probability $C_{il}$).
Then, the probability $P_{il}$ can be expressed as a sum of radiative and
collisional probabilities, $P_{il}=R_{il}+C_{il}$.
If these probabilities are multiplied with the starting level
population, then we call them radiative and collisional rates,
respectively.

For the case of a stationary atmosphere, i.e. if the time variations of
the medium are negligible, we may neglect the time derivative in the
equation \eqref{jk3eset} and we obtain the time independent set of the
statistical equilibrium equations,
\begin{equation}
\nabla \cdot \left(n_i \vec{v} \right)
= \sum_{l\ne i} \left(n_l P_{li} - n_i P_{il} \right).
\end{equation}
This set of equations has to be used in stationary moving atmospheres,
which includes the common case of a stellar wind.
However, due to simplicity, the advective term is very often neglected.
As a result we obtain an equation, which is valid in static atmospheres
exactly,
\begin{equation}\label{jk3ese}
\sum_{l\ne i} \left(n_l P_{li} - n_i P_{il} \right)=0.
\end{equation}
We say that the particular transition $i\leftrightarrow l$ is in
detailed balance, if $n_l P_{li} =  n_i P_{il}$.
If this condition is fulfilled for all $i,l$, then all transitions are
in detailed balance, which is the case of LTE.
Then, also the level populations have their equilibrium values
$n_i^\ast$ given by Eq.\,\eqref{jk3boltz}.

%=======================================================================
\subsection{Equilibrium level populations}

Equilibrium level populations are calculated using the assumption of
thermodynamic equilibrium or local thermodynamic equilibrium.
They are usually denoted as $n_i^\ast$ and referred to as LTE
populations.
However, \cite{mih} uses the quantity $n_i^\ast$ in a different meaning.
The LTE populations are calculated \emph{with respect to the ground
level of the next higher ion} using the equation \citep[Eq.5-14
in][]{mih}
\begin{equation}\label{jk3deflte}
n^*_{i,j} = n_{0,j+1} n_e \frac{g_{ij}}{g_{0,j+1}}
{\pul \zav{\frac{h^2}{2\pi m_e kT}}^\japul}
e^{-\frac{\chi_{Ij}-\chi_{ij}}{kT}}.
\end{equation}
Here $j$ is the index of the corresponding ion and $n_{0,j+1}$ is the
\emph{actual} population of the ground level of the next higher ion,
which is not necessarily an LTE value.
Equation \eqref{jk3deflte} is used as a definition of LTE populations in
NLTE calculations.
Then, if we are using the departure coefficients
$b_i= {n_i}/{n_i^\ast}$ \citep{menz}, it is necessary to know which
definition of LTE population is used.
Note that for both definitions of LTE populations the equation
\eqref{jk3boltz} holds and that for the case of LTE, $b_i=1$.

%=======================================================================
\subsection{Radiative rates}

The expressions for radiative rates shown here closely follow the book
of \cite{mih}.

%-----------------------------------------------------------------------
\subsubsection{Photoionization and photorecombination transitions}

\paragraph{Photoionization:}
The radiative rate for a photoionization transition is obtained by
expressing the
amount of absorbed energy $E$ during photoionization from the bound
state $i$ to the free state $k$, $E=4\pi J_\nu \alpha_{ik}(\nu) \dnu$.
The number of photoionizations is obtained by dividing the energy $E$ by
$h\nu$ and integrating from the ionization frequency $\nu_0$ to
$\infty$,
\begin{equation}\label{jk3rikup}
n_i R_{ik} = n_i 4 \pi \int_{\nu_0}^\infty \frac{\alpha_{ik}}{h\nu}
J_\nu \dnu
\end{equation}

\paragraph{Photorecombination:}
Photorecombination is a collisional process, since the ion catches the
free electron during this process.
If the medium is in a thermodynamic equilibrium, then from the detailed
balance it follows that $J_\nu=B_\nu$.
The number of recombinations has its equilibrium value
\begin{equation}
\zav{n_i R_{ik}}^\ast =
\zav{n_k R_{ki}}^\ast = n_i^\ast 4 \pi \int_{\nu_0}^\infty
\frac{\alpha_{ik}(\nu)}{h\nu} B_\nu \dnu.
\end{equation}
By adding and subtracting the expression $B_\nu e^{-\frac{h\nu}{kT}}$
under the integral and using the expression for the Planck function
\eqref{jk3planck}, we
obtain
\begin{equation}\label{jk3rekb}
\zav{n_k R_{ki}}^\ast = n_i^\ast 4 \pi \int_{\nu_0}^\infty
\frac{\alpha_{ik}(\nu)}{h\nu}
\hzav{\frac{2h\nu^3}{c^2} + B_\nu} e^{-\frac{h\nu}{kT}} \dnu.
\end{equation}
Dividing \eqref{jk3rekb} by $n_k^\ast$ yields an expression for a rate
per ion (the probability).
This expression has to be valid also outside LTE, so we replace the
Planck function $B_\nu$ by actual mean radiation intensity $J_\nu$.
Finally, multiplying the result by actual population of the free state
$n_k$ we obtain
\begin{equation}\label{jk3rikdown}
n_k R_{ki} =
n_k \zav{\frac{n_i^\ast}{n_k^\ast}}
4\pi
\int_{\nu_0}^\infty \frac{\alpha_{ik}(\nu)}{h\nu}
\hzav{\frac{2h\nu^3}{c^2}+J_\nu} e^{-\frac{h\nu}{kT}} \dnu
\end{equation}
where $n_i^\ast/n_k^\ast=n_e \Phi_{ik}(T)$ is the so called
Saha-Boltzmann factor.

%-----------------------------------------------------------------------
\subsubsection{Bound-bound transitions}

We consider transitions from the level $i$ to the level $l$.
The number of transitions caused by intensity $J(\nu)$ in the interval
$\dnu\domega$ is $n_i B_{il} \phi_{il}(\nu) J_\nu \dnu$,
where $\phi_{il}(\nu)$ is the line profile function for the given
transition, and $B_{il}$ is the Einstein coefficient for absorption.
Since $\alpha_{il}(\nu)= \alpha_{il}^0 \phi(\nu) =
\frac{h\nu}{4\pi} B_{il}\phi_{il}(\nu)$, we may
express the total number of absorptions obtained by integration over the
line profile as
\begin{align}\label{jk3rijup}
n_i R_{il} = n_i B_{il} \int \phi_{il}(\nu) J_\nu \dnu & =
n_i 4\pi \int \frac{\alpha_{il}(\nu)}{h\nu} J_\nu \dnu.
\end{align}
The total number of emissions is given by a sum
of spontaneous and stimulated emissions
\begin{equation}
n_l R_{li} = n_l \zav{A_{li} + B_{li}
\int\phi_{il}(\nu) J_\nu \dnu}
= n_l \frac{4\pi}{h\nu_{il}}
\frac{g_i}{g_l} \alpha_{il}^0 \hzav{\frac{2h\nu_{il}^3}{c^2} +
\int\frac{\phi_{il}(\nu)}{h\nu} J_\nu \dnu}.
\end{equation}
Using the Boltzmann equation \eqref{jk3boltz}, we may write the total
number of de-excita\-tions as
\begin{equation}\label{jk3rijdown}
n_l R_{li} = n_l \frac{n_i^\ast}{n_l^\ast} 4\pi
\int \frac{\alpha_{il}(\nu)}{h\nu}
\zav{\frac{2h\nu^3}{c^2} + J_\nu} e^{-\frac{h\nu}{kT}} \dnu.
\end{equation}

%-----------------------------------------------------------------------
\subsubsection{Net radiative bracket}
The net rate for the transition $i\leftrightarrow l$ may be obtained by
subtracting upward and downward radiative rates
\begin{align}
n_l \zav{A_{li}+B_{li} \int\phi_{il}(\nu) J_\nu \dnu} -
n_i B_{il} \int\phi_{il}(\nu) J_\nu \dnu \equiv n_j A_{li} Z_{li}.
\end{align}
We have introduced the net radiative bracket $Z_{li}$ as
\begin{align}
Z_{li} = 1- \frac{n_i B_{il} - n_j B_{li}}{n_l A_{li}}
\int\phi_{il}(\nu) J_\nu \dnu
= 1-\frac{\int\phi_{il}(\nu) J_\nu \dnu}{S_{il}},
\end{align}
where $S_{il}$ is the source function (independent of $\nu$) of the
transition $i\leftrightarrow l$.
For the case of detailed radiative balance $Z_{li}=0$.

%=======================================================================
\subsection{Collisional rates}

Inelastic collisions transfer energy between the thermal
particle motion energy and the internal energy of atoms.
Although collisions between all particles occur in stellar atmospheres,
it is mostly sufficient to consider only collisions with  electrons,
because the ratio between electron and ion thermal velocities is
$v_\mathrm{th,e}/v_\mathrm{th,i} \approx 43 \sqrt{A}$, consequently, the
ion contribution to collisions can be neglected.

The upward collisional rate from the level $i$ to the level $l$ is
obtained by integration of the total cross section of the transition
$\sigma_{il}(v)$ over the electron velocity distribution $f(v)$, which
is Maxwellian in our case,
\begin{equation}\label{jk3cup}
n_i C_{il} = n_i n_e \int_{v_0}^\infty \sigma_{il}(v) f(v) v \der v
= n_i n_e q_{il}(T),
\end{equation}
where the quantity $q_{il}(T)$ is called the effective collision
strength \citep[see, e.g.][Eq.\,5-68]{mih}.
The downward rate follows from the detailed balance in thermodynamic
equilibrium, $n_l^\ast C_{li} = n_i^\ast C_{il}$, and is
\begin{equation}\label{jk3cdown}
n_l C_{li} = n_l \zav{\frac{n_i^\ast}{n_l^\ast}} C_{il}
= n_l \zav{\frac{n_i^\ast}{n_l^\ast}} n_e q_{il}(T).
\end{equation}
Thus, for calculation of both upward and downward collisional rates we
need to know only the effective collision strength $q_{il}(T)$.

Details of calculations of collisional rates can be found elsewhere in
these proceedings \citep{nicekb,nicelm}.

%=======================================================================
\subsection{Set of statistical equilibrium equations}

Let us consider an element having {\NL} levels.
For each level $i$ we may write an equation
\begin{equation}
n_i \sum_{\substack{l=1 \\ l\ne i}}^\NL \zav{R_{il}+C_{il}} +
\sum_{\substack{l\ne i \\ l=1}}^\NL n_l \zav{R_{li}+C_{li}} = 0,
\end{equation}
where $R_{il}$ and $C_{il}$ are given by equations
\eqref{jk3rikup},
\eqref{jk3rikdown},
\eqref{jk3rijup},
\eqref{jk3rijdown},
\eqref{jk3cup}, and
\eqref{jk3cdown}.
Since the system of these equations for all levels is linearly dependent
(for {\NL} levels only $\NL-1$ levels are linearly independent), we
must supplement the equations by some additional equation.

For model atmosphere calculations, we have two equivalent options (both
physically and numerically).
We can use either the charge conservation equation,
\begin{equation}
\sum_k \sum_j j N_{jk} = n_e,
\end{equation}
where $N_{jk}$ means the number density of ions $j$ of the element $k$,
or the particle conservation,
\begin{equation}
\sum_k \sum_j N_{jk} = (N-n_e),
\end{equation}
where $N$ is the total number density in the atmosphere,
are being used.

For the case of the solution of the NLTE problem for the trace element,
the abundance equation
\begin{equation}\label{jk3abeq}
\sum_j N_{jk} = \frac{\alpha_k}{\alpha_\mathrm{ref}} \sum_j
N_{j,\mathrm{ref}}
\end{equation}
can be used.
Here $\alpha_k$ is the abundance of the element $k$ and
$\alpha_\mathrm{ref}$
is the abundance of the reference atom, which is usually (but not
necessarily) hydrogen.
Equation \eqref{jk3abeq} relates the abundances of the studied and
reference elements.

%=======================================================================
\subsection{Non-explicit levels}

Usually the atoms that exist in reality have much more complicated
structure than we are able to include in our models.
As a consequence, some energy levels are not explicitly considered in
the statistical equilibrium equations.
This mainly concerns highly excited levels.
However, it does not mean that they do not exist and we have to include
them in the system of the statistical equilibrium equations somehow.
There are several possibilities of how to handle this task.
We may, of course, neglect them.
However, it is better to include them in the total number density and to
assume that they are in LTE with respect to the next higher ion.
Also the corresponding collisional excitation rates to these levels can
then be included in the generalised (or modified) collisional ionization
rate \citep[for a detailed description of this approximation see][]
{tlustypopis,hhl}.
The radiative rates to these levels are assumed to be in detailed
radiative balance and they do not enter into the equations of
statistical equilibrium.

Alternatively, they may be merged into one or more superlevels
\citep[see][]{nltebl1}, which can then be included in the set of energy
levels and in the equations of statistical equilibrium to allow their
departures from the local thermodynamic equilibrium.

%%%%%%%%%%%%%%%%%%%%%%%%%%%%%%%%%%%%%%%%%%%%%%%%%%%%%%%%%%%%%%%%%%%%%%%%
\section{Conclusions}

In this review we summarized the basic equations for the solution of the
NLTE line formation problem with a focus on the solution of the problem
for trace elements in stellar atmospheres.
It has to be emphasized that in NLTE calculations for trace elements
preference should be given to background NLTE model atmospheres, since
using LTE model atmospheres is inconsistent with the condition of NLTE.
It must be carefully decided which elements can be treated as a trace
element and for which such treatment is questionable.
After obtaining the numerical results all assumptions have to be
verified, especially that of the traceness.

\begin{acknowledgements}
The author thanks Drs. Barry Smalley and Glenn Wahlgren for their
comments to the manuscript.
This work was partly supported by a GA \v{C}R grant 205/07/0031.
The Astronomical Institute Ond\v{r}ejov is supported by a project
AV0Z10030501.
\end{acknowledgements}

\newcommand{\nicesbornik}[1]{in Non-LTE Line Formation for Trace
	Elements in Stellar Atmospheres, R. Monier et al. eds., EAS
	Publ. Ser. Vol. x, p.~#1}
\newcommand{\rokvydani}{2010}

\end{document}